\documentstyle[twocolumn,aps,epsf]{revtex}

\title{Dynamics of liquid $^4$He in Vycor}

\author{H.R. Glyde}
\address{Department of Physics and Astronomy, University of Delaware,
Newark, Delaware 19716}

\author{O. Plantevin and B. F\aa k}
\address{D\'{e}partement de Recherche Fondamentale sur la Mati\`{e}re
Condens\'{e}e, SPSMS/MDN, CEA Grenoble, 38054 Grenoble, France}

\author{G. Coddens}
\address{Laboratoire L\'{e}on Brillouin, CEA/CNRS,
91191 Gif-sur-Yvette Cedex, France}

\author{P.S. Danielson}
\address{Corning Incorporated, Sullivan Park FR-5, Corning,
New York 14831}

\author{H. Schober}
\address{Institut Laue Langevin, BP 156, 38042 Grenoble, France}

\date{September 12, 1999}

\begin{document}
\maketitle

\begin{abstract}
We have measured the dynamic structure factor of liquid $^4$He in
Vycor using neutron inelastic scattering.  Well-defined phonon-roton
(p-r) excitations are observed in the superfluid phase
for all wave vectors $0.3 \leq Q \leq
2.15$ \AA$^{-1}$.  The p-r energies and lifetimes at low temperature
$(T = 0.5$ K) and their temperature dependence are the same as in bulk
liquid $^4$He.  However, the weight of the single p-r component does
not scale with the superfluid fraction $\rho_S(T)/\rho$ as it does in
the bulk.  In particular, we observe a p-r excitation between
$T_c=1.952$ K, where $\rho_s(T) = 0$, and $T_\lambda=2.172$~K of the
bulk.  This suggests, if the p-r excitation intensity scales with the
Bose condensate, that there is a separation of the Bose-Einstein
condensation temperature and the superfluid transition temperature
$T_c$ of $^4$He in Vycor.  We also observe a two-dimensional layer
mode near the roton wave vector.  Its dispersion is consistent with
specific heat and $\rho_S(T)$ measurements and with layer modes
observed on graphite surfaces.
\end{abstract}
\pacs{PACS No. 67.40.-w}

Liquid $^4$He immersed in Vycor is a readily accessible example of
bosons in disorder and confinement.  The specific heat, superfluid
density and other thermodynamic properties of this ``dirty Bose
system'' have been extensively investigated \cite{reppy92,brewer78} to
reveal the impact of disorder and finite length scales on excitations
and phase transitions \cite{thouless74}.  Understanding gained in
helium can be transferred to other examples of bosons in disorder such
as flux lines in superconductors \cite{supracon}, granular metal films
\cite{granular}, Cooper pairs in Josephson junction arrays
\cite{supracon} and (possibly) Cooper pairs in high-$T_c$ materials if
pairing occurs above the Bose-Einstein condensation (BEC) temperature
that leads to superconductivity \cite{randeria95}.  Direct measurement
\cite{coddens,sokol,dimeo98,OP,rain99,azuah99} and simulation
\cite{makivic93,krauth91,boninsegni98} of excitations of liquid $^4$He
in disorder, however, have only recently begun.  In this Letter, we
present neutron scattering measurements of the dynamic structure
factor $S(Q,\omega)$ of liquid $^4$He in 30\% porous Vycor over a wide
wave-vector range, $0.3 \leq Q \leq 2.15$ \AA$^{-1}$, and temperature
range, $0.5 \leq T \leq 2.31$ K. We present the first observation of
phonons in Vycor, evidence for a two-dimensional (2D) layer mode, and
evidence that the superfluid transition temperature $T_c$ and the BEC
temperature of liquid $^4$He in Vycor may not be the same.

The present Vycor sample, a cylinder of 9.7 mm diameter and 40 mm
height, was synthesized in the usual way except that natural boron was
replaced by $^{11}$B (99.95\% purity). Natural boron has
a large absorption cross-section for neutrons while $^{11}$B does not.
Small-angle neutron scattering measurements on the present sample
showed that it has the same static structure factor $S(Q)$ as standard
Vycor plates made with natural boron \cite{drycor}. The sample was
fitted into a tightly machined cylindrical aluminum sample holder of
1.5 mm wall thickness and 100 mm height.  The sample was fully filled
with helium with a compartment of bulk $^4$He above the Vycor for
reference measurements.  The Vycor and bulk $^4$He compartments were
separated by a Cd spacer. The sample cell was mounted in a $^3$He
cryostat, where the thermometers were calibrated against the $^3$He
vapor pressure in the sample cell in a separate run. The temperature
was regulated within $\pm 0.02$ K. The measurements
were made on the IN6 time-of-flight spectrometer at the Institut Laue
Langevin, using an incident neutron energy of 3.83 meV and an energy
resolution (FWHM) of about 110 $\mu$eV.

Figure 1(a) shows $S(Q,\omega)$ of liquid $^4$He in Vycor for a wave
vector $Q$ in the phonon region, $Q = 0.35$ \AA$^{-1}$.  At $T = 0.5$
K, $S(Q,\omega)$ is confined almost entirely to a single peak arising
from creation of single phonons in the liquid by the neutrons.
Multiphonon creation is small at these wave vectors.  The peak width
is set by the instrument resolution width.
As $T$ is increased to $T = 2.31$ K, the single phonon peak broadens,
but there is still a well defined peak in the normal phase.  This
shows that liquid $^4$He in Vycor supports a well defined sound mode
in both the superfluid $(T = 0.5$ K) and normal $(T = 2.31$ K) phase,
as in bulk liquid $^4$He.
This is the first direct observation of phonons in Vycor by neutron
scattering.

Figure 1(b) shows $S(Q,\omega)$ at several temperatures for $Q = 1.7$
\AA$^{-1}$, a wave vector between the maxon $(Q = 1.1$ \AA$^{-1})$
and the roton $(Q = 1.95$ \AA$^{-1})$ regions.  There is a well defined,
single phonon-roton (p-r) excitation in $S(Q,\omega)$ at the lowest
temperature, $T = 0.5$ K. As $T$ increases, the p-r peak broadens and
the integrated intensity in the peak decreases.  At $T = 2.31$ K,
there is no discernible peak in $S(Q,\omega)$.  However, at $T = 1.99$
K, which is above $T_c = 1.952$ K \cite{Chan88} (where $\rho_S = 0$) 
but below the
bulk value for $T_\lambda=2.172$ K, there is still a thermally
broadened peak.  Woods and Svensson \cite{woods78} proposed that for
wave vectors at the maxon and higher $(Q \agt 1.1$ \AA$^{-1}$), the
weight of the characteristic maxon-roton excitation in $S(Q,\omega)$
of bulk liquid $^4$He scaled as the superfluid fraction,
$\rho_S(T)/\rho$.  There is no mode for $Q \agt 1.0$ \AA$^{-1}$ in
normal bulk \cite{glyde93,griffin93} liquid $^4$He where $\rho_S(T) =
0$.  We return to this point at the end of this Letter.

Figure 2(a) shows $S(Q,\omega)$ at the roton wave vector $Q = 1.95$
\AA$^{-1}$ and $T = 0.5$ K. The intense peak at $\omega=0.74$ meV
arises from exciting the p-r mode at the roton wave vector.  We call
this the 3D roton.  The p-r energies $\omega_Q$ for $0.3 \leq Q \leq
2.15$ \AA$^{-1}$, obtained from $S(Q,\omega)$ at many $Q$ values such
as shown in Figs.\ 1 and 2(a), are displayed in Fig.\ 3.  The phonon-roton
energies in Vycor are the same as in the bulk
(perhaps marginally lower for $Q \agt 1.9$ \AA$^{-1}$), within the
present experimental precision $(\pm 5\: \mu$eV).

Returning to Fig.\ 2(a), we see that there is additional intensity at
energies below the 3D roton peak in Vycor that is not seen in bulk
liquid $^4$He \cite{glyde93,griffin93}.  This additional intensity is
small, approximately 8\% of the main 3D \mbox{p-r} integrated intensity at
the roton.  We emphasize that $S(Q,\omega)$ of liquid $^4$He in both
Vycor and bulk shown in Fig.\ 2(a) are from the present measurements.
We observe the additional intensity in Vycor for $Q \agt 1.7$
\AA$^{-1}$ only.  The intensity of the new mode is shown in Fig.\ 2(b)
with its energy dispersion as an inset.  For $Q \alt 1.7$ \AA$^{-1}$,
the intensity in the new mode either becomes too weak to be observed,
or the mode energy lies sufficiently close to the 3D p-r mode that it
cannot be resolved from the p-r peak.

Since all measurements were made on fully filled Vycor, we cannot
identify the region of the liquid from which the additional scattering
originates.  However, we interpret the additional intensity as a 2D
layer mode propagating in the liquid layers adjacent to the two solid
$^4$He layers on the Vycor surfaces.  The layer mode energy near its
``roton'' minimum is well described by $\omega(Q) = \Delta_{2D} +
(Q-Q_{2D})^2/2\mu$ with $\Delta_{2D} = 0.55 \pm 0.01$ meV, $Q_{2D} =
1.94 \pm 0.01$ \AA$^{-1}$ and $\mu = 0.13 \pm 0.01\ m_4$ ($m_4$ is the
$^4$He atomic mass).  The present ``gap energy'' $\Delta_{2D}$ is
consistent with the gap energy $0.54 \pm 0.03$ meV of 2D rotons on
graphon surfaces observed by Thomlinson {\it et
al.} \cite{thomlinson80} and with that of 0.6 meV observed by Lauter {\it et
al.} \cite{lauter92} on graphite surfaces.
The difference between the 3D roton energy
$\Delta = 0.742$ meV and $\Delta_{2D}$ is consistent with the
differences predicted originally by Padmore \cite{padmore74} for 2D
rotons and calculated more recently by Clements {\it et
al.} \cite{clements96}. The calculations by Clements {\it et al.} also
suggest that the layer mode intensity is small at lower $Q$ as found
here.  The present gap energy $\Delta_{2D} = 0.55 \pm 0.01$ meV is
consistent with the gap energy of 0.53 meV obtained by Brewer {\it et
al.} \cite{brewer65} for the layer mode contribution to the specific
heat in Vycor.  Kiewiet {\it et al.} \cite{kiewiet75} found that the
superfluid density $\rho_S(T)$ in Vycor for $T \leq 1.4$ K is well
described if the normal density arises from exciting ``one-dimensional''
phonons and a roton-like mode having a roton gap of 0.50
meV. This interpretation is consistent with the phonons observed here,
which will propagate predominantly along the pores, and with the
present 2D layer mode.  For $T \leq 1.4$ K, the 3D roton energy
$\Delta=8.62$ K is too high for 3D rotons to be excited.
Thus, the interpretation of Kiewiet {\it et al.} \cite{kiewiet75} is
consistent with the phonons and the 2D layer mode that we observe
here.
The additional intensity shown in Fig.\ 2(a) is similar to that
observed by Dimeo {\it et al.} \cite{dimeo98} at an energy of 0.3-0.5
meV at the roton wave vector, although their intensity is two times greater
(20\% of the 3D roton). However, they could not determine the mode energy
with precision.

To determine how the weight of the p-r mode component in $S(Q,\omega)$
scales with temperature, we have
fitted the following model to the data,
\begin{equation}
\chi''(Q,\omega) = f_S(T) \chi_S''(Q,\omega) + f_N(T)\chi_N''(Q,\omega),
\label{WS}
\end{equation}
where $S(Q,\omega)=[1-\exp(-\hbar\omega/k_BT)]^{-1}\chi''(Q,\omega)$.
At $T = 0.5$ K, $\chi''_S(Q,\omega)$ is the total observed $\chi''(Q,\omega)$,
corresponding predominantly to the single p-r peak [see Fig.
1(b)].  $\chi''_N(Q,\omega)$ is the total $\chi''(Q,\omega)$ observed
at $T = 2.31$ K, which contains no p-r peak at all.  The $f_S(T)$ is a
free parameter that we obtain by fitting Eq.\ (\ref{WS}) to
$\chi''(Q,\omega)$ observed at temperatures between $T = 0.5$ and 2.31
K, and $f_N(T) = 1-f_S(T)$.  In the fit the p-r mode energy and width
in $\chi''_S(Q,\omega)$ was allowed to vary with temperature.
Clearly, the model requires $f_S(T) = 1$ at $T = 0.5$ K and $f_S(T) =
0$ at $T = 2.31$ K.  The fitted fraction $f_S(T)$ represents,
approximately, the fraction of the total $S(Q,\omega)$ taken up by the
single excitation peak.

Figure 4 shows the fitted values of $f_S(T)$ compared with the
superfluid fraction $\rho_S(T)/\rho$ in Vycor and in bulk liquid
$^4$He.  The fraction $f_S(T)$ in Vycor clearly does not scale with
$\rho_S(T)/\rho$ in Vycor.  At low temperatures,  both $f_S(T)$ and
$\rho_S/\rho$ are approximately unity, while at higher $T$, $f_S(T)$
is still large whereas $\rho_S(T)/\rho\approx 0$.  In searching for
possible causes, we note that there is some bulk liquid $^4$He between
the Vycor sample and the sample cell walls.  The fraction of such bulk
liquid to liquid in Vycor pores is estimated to be at most 10\%.  This
small fraction could not account for the large deviation of $f_S(T)$
from $\rho_S(T)/\rho$ in Vycor.  We would essentially need all of the
liquid to be bulk liquid to explain the scaling in Fig.\ 4.  Also,
since $\chi''_N(Q,\omega)$ is expected to be largely independent of $T$ for
$T \agt T_\lambda$ (as it is in bulk liquid $^4$He), the fitted
$f_S(T)$ should not be sensitive to the temperature at which
$\chi''_N(Q,\omega)$ is defined.

The central finding is therefore that the weight of the single
phonon-roton excitation peak in $S(Q,\omega)$ at higher $Q$ values
does not scale with $\rho_S(T)/\rho$ in Vycor.  The deviation is large
and there remains a p-r peak in $S(Q,\omega)$ above $T_c$ where
$\rho_S(T) = 0$.  Thus the apparent scaling of peak weight with
$\rho_S(T)$ in bulk $^4$He is not universal and does not extend to
confined geometries.  As noted, Glyde and Griffin (GG) \cite{GG90} proposed
that the sharp excitation in $S(Q,\omega)$ at $Q \agt 1$ \AA$^{-1}$
arises because there is a condensate and that the weight should scale
approximately as the condensate fraction, $n_0(T)$.  The excitation
weight in $S(Q,\omega)$ in Vycor might still scale with $n_0(T)$ (as
in bulk $^4$He) if $n_0(T)$ in Vycor were similar to bulk $^4$He and
particularly if $n_0(T)$ were finite between $T_c = 1.952$ K and
$T_\lambda = 2.172$ K. We therefore arrive at the interesting
conclusion: either the GG proposal is incorrect or there is a
(possibly localized)  condensate in $^4$He in Vycor between $T_c$ and
$T_\lambda$.  If the latter is true, then we are observing the
separation of the BEC temperature from the superfluid transition
temperature $(T_c)$ by disorder or confinement \cite{supracon,Huang}.
This intriguing
possibility needs to be clarified by further measurements of
$S(Q,\omega)$ between $T_c$ and $T_\lambda$ and direct measurements of
$n_0$.

We thank the staff at the Institut Laue Langevin for technical
assistance.  This work was partially supported by the National Science
Foundation, grant DMR 96-23961.

\onecolumn

\begin{figure}
\begin{center}
   \begin{minipage}[t]{1\textwidth}
    \centerline{\epsfxsize=.6\textwidth \epsffile{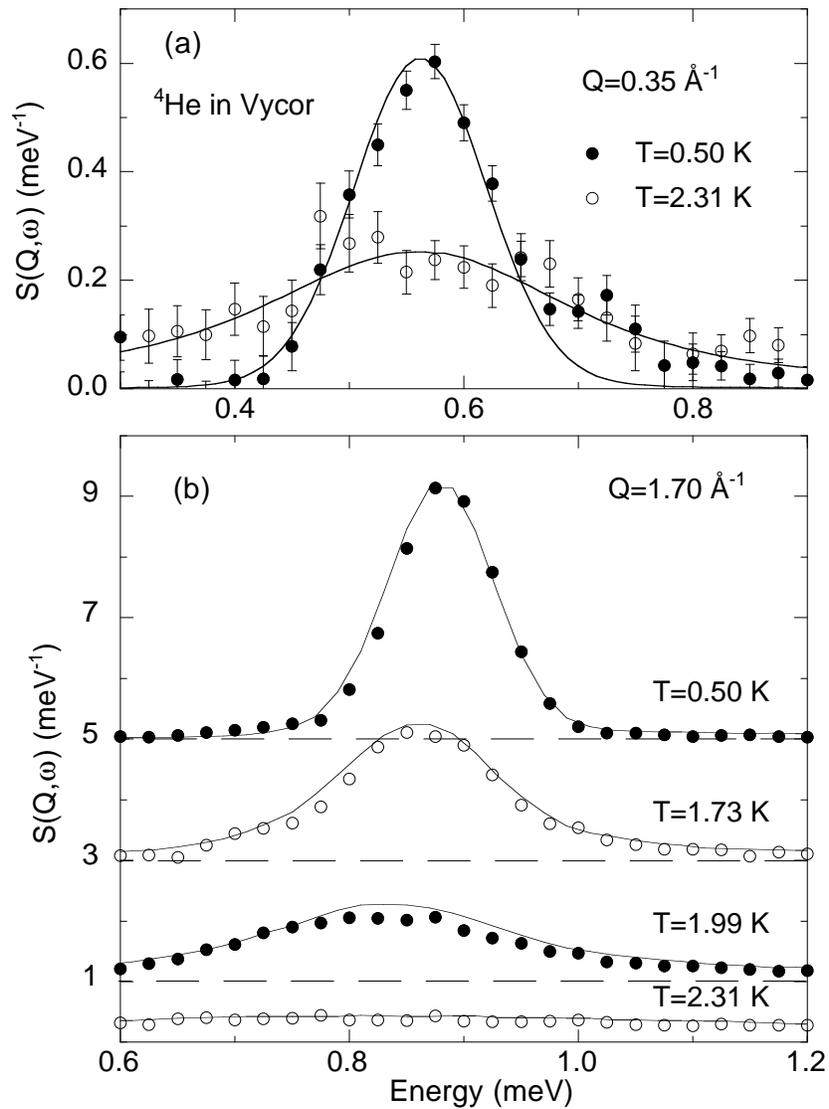}}
\caption{ \large
Dynamic structure factor of liquid $^4$He in Vycor at
temperatures as shown.
(a) $Q = 0.35$ \AA$^{-1}$ (phonon region).  The lines are guides
to the eye.
(b) $Q = 1.7$ \AA$^{-1}$ (between the maxon and roton regions).
Symbols are $^4$He in Vycor, with errors smaller than the symbol size.
The lines show bulk data for comparison.}
\label{Fig1}
   \end{minipage}
\end{center}
\end{figure} 

\newpage

\begin{figure}
\begin{center}
   \begin{minipage}[t]{1\textwidth}
    \centerline{\epsfysize=0.6\textheight \epsfbox[107 159 441 733]{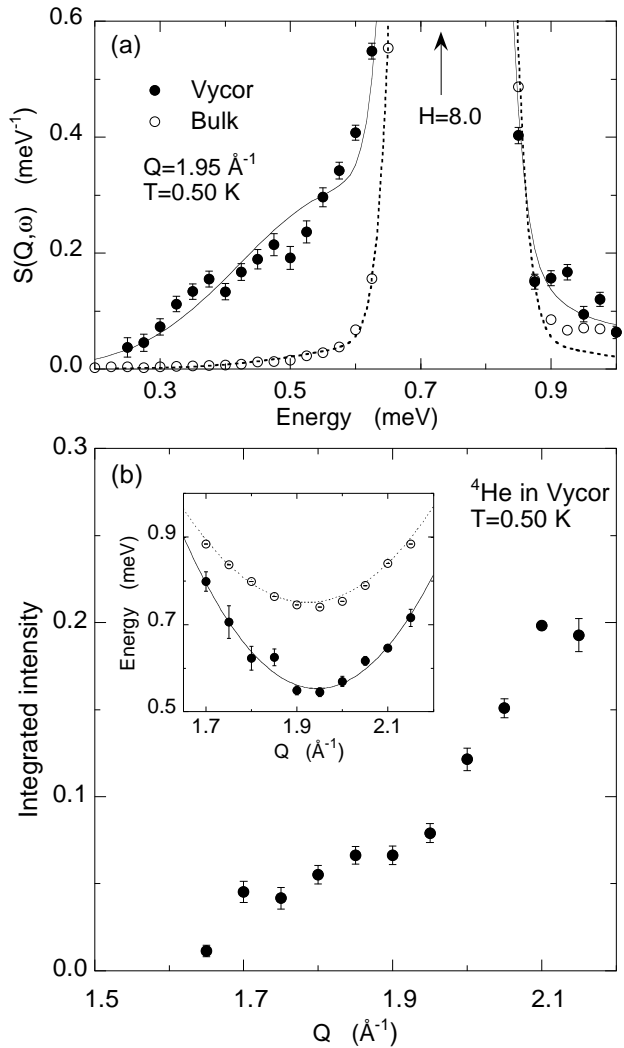}}
    \caption{ \large
(a) $S(Q,\omega)$ of liquid $^4$He at $T = 0.5$ K in Vycor
at the roton wave vector (solid circles and line).  The dashed line is
the component of $S(Q,\omega)$ arising from exciting the 3D roton in
the liquid (peak height $H = 8.0$ meV$^{-1}$).  The intensity at
energies below the roton peak is attributed to a 2D layer mode
propagating in the liquid layers adjacent to the Vycor walls.  The
open circles are the corresponding $S(Q,\omega)$ observed in bulk
$^4$He (present measurements).  (b)~Integrated intensity in the 2D
layer mode versus wave vector.  At $Q
= 1.95$ \AA$^{-1}$, the integrated intensity of the 2D mode is 8\% of
the 3D roton intensity in fully filled Vycor.  The inset shows the
energies of the 2D layer mode (solid circles and line) and the
3D roton (open circles) in Vycor as well as the 3D bulk roton energy
(dotted line).}
   \end{minipage}
 \end{center}
\label{Fig2}
\end{figure}

\newpage

\begin{figure}
\begin{center}
  \begin{minipage}[t]{1\textwidth}
    \centerline{\epsfxsize=0.6\textwidth \epsffile{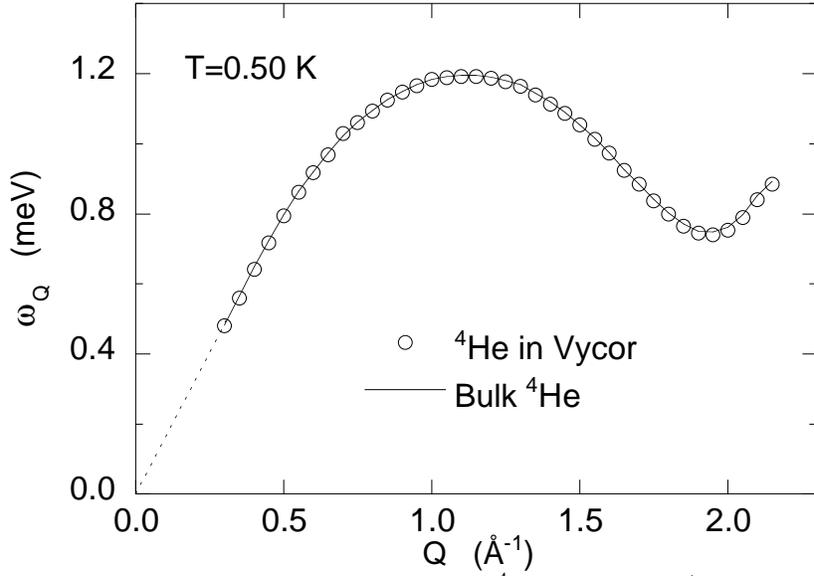}}
\caption{\large
	Phonon-roton energy dispersion curve of liquid $^4$He
in Vycor (open circles) and in bulk $^4$He (line).  The error
bars are much smaller than the symbol size.
	}
  \end{minipage}
\end{center}
\label{Fig3}
\end{figure}

\vspace{3cm}

\begin{figure}
\begin{center}
   \begin{minipage}[t]{1\textwidth}
    \centerline{\epsfxsize=0.6\textwidth \epsffile{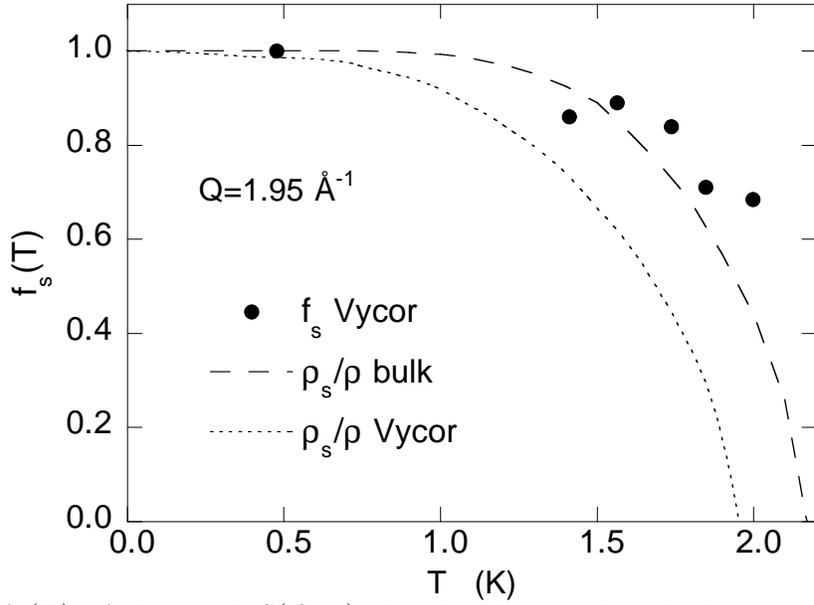}}
\caption{ \large
Fraction $f_S(T)$ of the total $S(Q,\omega)$ that is taken
up by the low-temperature component $S_S(Q,\omega)$ (chiefly the
single-excitation component) as a function of temperature for Vycor
[see Eq.\ (\protect\ref{WS})].  The $f_S(T)$ in Vycor does not scale
with the superfluid fraction $\rho_S(T)/\rho$ of Vycor (dotted line;
Ref.\ \protect\cite{reppy92}).  The dashed line is $\rho_S(T)/\rho$
for bulk $^4$He \protect\cite{reppy92}.}
   \end{minipage}
\end{center}
\label{Fig4}
\end{figure}
\end{document}